\documentclass{iaus}
\usepackage{graphicx}

\def\al{\alpha}

\def\ga{\gamma}

\def\ka{\kappa}
\def\la{\lambda}

\def\ps{\psi}

\def\De{\Delta}

\def\cL{{\cal L}}

\def\mn{{\mu\nu}}

\def\fr#1#2{{{#1} \over {#2}}}
\def\half{{\textstyle{1\over 2}}}
\def\quar{{\textstyle{1\over 4}}}
\def\frac#1#2{{\textstyle{{#1}\over {#2}}}}

\def\ol#1{\overline{#1}}

\def\pt#1{\phantom{#1}}

\def\sb{\overline{s}}

\def\ab{\overline{a}}
\def\atw{\widetilde{a}}

\def\hul#1#2{h^{#1}_{{\pt{#1}}#2}}

\def\vb#1#2{e_{#1}^{{\pt{#1}}#2}}
\def\ivb#1#2{e^{#1}_{{\pt{#1}}#2}}
\def\uvb#1#2{e^{#1#2}}

\def\mt{m^{\rm T}}
\def\ms{m^{\rm S}}

\def\oaut{{\ol a}^{\rm T}}
\def\oaus{{\ol a}^{\rm S}}

\def\lrDmu{{\hskip -3 pt}\stackrel{\leftrightarrow}{D_\mu}{\hskip -2pt}}

\newcommand{\beq}{\begin{equation}}
\newcommand{\eeq}{\end{equation}}
\newcommand{\bea}{\begin{eqnarray}}
\newcommand{\eea}{\end{eqnarray}}
\newcommand{\rf}[1]{(\ref{#1})}

\title[Lorentz Violation and Gravity]{Lorentz Violation and Gravity}

\author[Quentin G.\ Bailey]{Quentin G.\ Bailey}

\affiliation{Physics Department,
Embry-Riddle Aeronautical University,\\
3700 Willow Creek Road,
Prescott, AZ 86301, USA,\\
email: {\tt baileyq@erau.edu}}

\pubyear{2009}
\volume{261}  
\pagerange{1--5}
\date{May 2009}
\jname{Relativity in Fundamental Astronomy: Dynamics, 
Reference Frames, and Data Analysis}
\editors{S.\ Klioner, P.K.\ Seidelmann \& M.\ Soffel, eds.}
\begin{document}

\maketitle

\begin{abstract}
In the last decade, 
a variety of high-precision experiments have searched 
for miniscule violations of Lorentz symmetry.  
These searches are largely motivated by the possibility 
of uncovering experimental signatures 
from a fundamental unified theory.  
Experimental results are reported in the framework 
called the Standard-Model Extension (SME), 
which describes general Lorentz violation 
for each particle species 
in terms of its coefficients for Lorentz violation.  
Recently, 
the role of gravitational experiments 
in probing the SME has been explored in the literature.  
In this talk, 
I will summarize theoretical and experimental aspects 
of these works.  
I will also discuss recent lunar laser ranging 
and atom interferometer experiments, 
which place stringent constraints on gravity coefficients 
for Lorentz violation. 

\end{abstract}

\firstsection 
\section{Introduction}
 
General Relativity (GR) encompasses all known 
gravitational phenomena and remains the best known 
fundamental theory of gravity.  
No convincing deviations from GR have been detected 
thus far (\cite[Will 2005]{cw06}, 
\cite[Battat et al.\ 2007]{battat07},
\cite[M\"uller et al.\ 2008]{hm08},
and \cite[Chung et al.\ 2009]{c09}).
Nonetheless, 
there remains widespread interest in
continuing ever more stringent tests of GR 
in order to find possible deviations.
This is motivated by the idea that
small deviations from GR might be the 
signature of an as yet unknown
unified fundamental theory that successfully meshes 
GR with the Standard Model of particle physics.

The principle of local Lorentz symmetry
is a solid foundation of GR.
However, 
the literature abounds with 
theoretical scenarios in which this 
symmetry principle might be broken.
One promising possibility is that
miniscule deviations from perfect local
Lorentz symmetry might be observable in
high-precision experiments and observations
(for reviews see \cite[CPT 2008]{cpt08},
\cite[Bluhm 2006]{rb06}, 
and \cite[Amelino-Camelia et al.\ 2005]{ac05}).
The Standard-Model Extension (SME)
is a general theoretical framework for tests of Lorentz symmetry, 
in both gravitational and nongravitational scenarios
(\cite[Kosteleck\'y \& Potting 1995]{kp95},
\cite[Colladay \& Kosteleck\'y 1997]{ck97},
\cite[Colladay \& Kosteleck\'y 1998]{ck98},
and \cite[Kosteleck\'y 2004]{k04}).
The SME is an effective field theory that 
includes all possible Lorentz-violating terms in the action
while also incorporating the known physics of the 
Standard Model and GR.
The Lorentz-violating terms are constructed from 
gravitational and matter fields and coefficients for Lorentz violation,
which control the degree of Lorentz-symmetry breaking.

A wide variety of experiments with matter
have set constraints on many of the coefficients
for Lorentz violation in the SME
(for a summary and references see the data tables 
in \cite[Kosteleck\'y \& Russell 2009]{kr09}).
Recently, 
the gravitational sector and matter-gravity couplings 
in the SME have been studied 
(\cite[Bailey \& Kosteleck\'y 2006]{bk06},
\cite[Bailey 2009]{b09}, and
\cite[Kosteleck\'y \& Tasson 2009]{kt09}).  
In turns out that some novel effects can occur that are
controlled by certain matter sector coefficients in the SME, 
which are unobservable in the absence of gravity.
In addition, 
experimental work constraining SME coefficients in the 
pure-gravity sector has begun, 
including gravimeter experiments 
(\cite [M\"uller et al.\ 2008]{hm08})
and lunar laser ranging (\cite[Battat et al.\ 2007]{b07}).
The theoretical and experimental aspects of the signals 
for Lorentz violation in gravitational experiments 
will be discussed in this talk.
For convenience, 
natural units ($\hbar=c=1$) are used.

\section{Theory}

In \cite[Kosteleck\'y (2004)]{k04},
the SME with both gravitational and nongravitational couplings 
was presented in the general context of a Riemann-Cartan spacetime. 
This extended earlier work on the Minkowski spacetime 
limit. 
We focus here on the special cases of 
gravity couplings in the matter sector of the SME
and also on the pure-gravity sector. 

In the matter sector of the SME, 
the matter-gravity couplings expected to
dominate in many experimental scenarios 
for ordinary matter (protons, neutrons, and electrons)
can be described in terms of Dirac spinor fields for
spin-1/2 fermions.
The lagrangian density for this limit takes the form
\beq
\cL_m = \frac 12 i e \ivb{\mu}{a} {\overline \ps} 
(\ga^a 
- c_{\nu\la} \uvb{\la}{a}\ivb{\nu}{b}\ga^b+...) 
\lrDmu \ps 
-e {\overline \ps} 
(m+a_\mu \ivb{\mu}{a} \ga^a +... ) \ps+...,
\label{matter}
\eeq
where the ellipses represent additional terms in the SME, 
neglected here for brevity.
Following standard methods, 
the spinor fields $\ps$ and the gamma matrices 
$\ga^a$ are incorporated into the tangent space
of the spacetime at each point using the vierbein $\vb{\mu}{a}$.
The symbol $e$ represents the determinant of the vierbein.
The coefficients for Lorentz violation appearing
in equation \rf{matter} are $c_\mn$ and $a_\mu$, 
which vanish in the limit of perfect local Lorentz symmetry
for the matter fields.
Note also that the covariant derivative appearing in this lagrangian
is both the spacetime covariant derivative and the $U(1)$ 
covariant derivative, 
and so contains additional couplings to gravity 
through the spacetime connection. 

In the pure-gravity sector of the SME, 
and in the Riemann-spacetime limit, 
the relevant lagrangian is written as
\bea
\cL_g &=& \fr {1}{2\ka} e[(1-u)R + s^\mn R^T_\mn 
+ t^{\ka\la\mu\nu} C_{\ka\la\mu\nu}]+ \cL^\prime.
\label{gravity}
\eea
In this expression, 
$R$ is the Ricci scalar, 
$R^T_\mn$ is the trace-free Ricci tensor, 
$C_{\ka\la\mu\nu}$ is the Weyl conformal tensor, 
and $\ka=8\pi G$, 
where $G$ is Newton's gravitational constant.  
The leading Lorentz-violating gravitational couplings
are controlled by the 20 coefficients for Lorentz violation 
$u$, 
$s^\mn$, 
and $t^{\ka\la\mu\nu}$.
The matter sector and possible dynamical terms
governing the $20$ coefficients are
contained in the additional term denoted $\cL^\prime$.
In the limit of perfect local Lorentz symmetry 
for gravity these coefficients vanish.

In all sectors, 
the action in the SME effective field theory 
maintains general coordinate invariance.
However, 
under local Lorentz transformations
and diffeomorphisms of the localized matter 
and gravitational fields, 
or what are called \it particle transformations, \rm
the action is not invariant.
When Lorentz violation is introduced in the context
of a general Riemann-Cartan geometry, 
some interesting geometric constraints arise.
For example, 
introducing the coefficients for Lorentz violation 
in the matter and gravity sectors as 
nondynamical or prescribed functions generally conflicts 
with the Bianchi identities.
If instead the coefficients arise through a dynamical process, 
as occurs in a spontaneous Lorentz-symmetry breaking scenario,
conflicts with the geometry 
are avoided (\cite[Kosteleck\'y 2004]{k04}).

The approach of \cite[Kosteleck\'y \& Tasson (2009)]{kt09}
and \cite[Bailey \& Kosteleck\'y (2006)]{bk06} is to treat 
the coefficients for Lorentz violation as arising from 
spontaneous Lorentz-symmetry breaking.
Although specific models with dynamical vector and
tensor fields can reproduce the terms in the lagrangians
\rf{matter} and \rf{gravity}, 
it is a challenging task to study the gravitational
effects in a generic, model-independent way. 
However, 
the weak-field or linearized gravity regime 
offers some simplifications to the analysis.
It is possible, 
under certain mild assumptions on the dynamics
of the coefficients for Lorentz violation,
to extract effective linearized equations. 
In this case the effects of Lorentz violation
on gravity and matter involve only the vacuum expectation
values of the coefficients for Lorentz violation
(denoted ${\overline a}_\mu$, $\sb^\mn$, etc.).

By including matter-gravity couplings, 
the ${\overline a}_\mu$ coefficients, 
which have remained largely elusive in 
nongravitational tests, 
become accessible to experiments. 
In contrast, 
stringent constraints on the $c_\mn$ coefficients already exist 
(see the data tables in \cite[Kosteleck\'y \& Russell 2009]{kr09}).
As shown in \cite[Kosteleck\'y \& Tasson (2009)]{kt09}, 
the dominant Lorentz-violating effects on matter fields arise 
from an effective vector potential $\atw_\mu$ 
that takes the form
\beq
\atw_\mu = 
\half \al h_\mn \ab^\nu - \quar \al \ab_\mu \hul{\nu}{\nu}, 
\label{aftw}
\eeq
where $h_\mn$ are 
the metric fluctuations around a
Minkowski background and $\al$ is a constant.

In the pure-gravity sector, 
the weak-field gravity analysis in 
\cite[Bailey \& Kosteleck\'y (2006)]{bk06} 
reveals that the leading Lorentz-violating effects are controlled
by the nine independent coefficients in $\sb^\mn$.
In the post-newtonian limit, 
attempting to match the pure-gravity sector of 
the SME to the standard Parametrized Post-Newtonian (PPN) 
formalism involves constraining $\sb^{\mn}$ to an isotropic form
in a special coordinate system 
with one independent coefficient $\sb^{00}$.
This reveals that there is 
a partially overlapping relationship
between the two approaches
(see figure 1 in \cite[Bailey \& Kosteleck\'y 2006]{bk06}). 
Thus, 
the pure-gravity sector of the SME describes 
effects outside of the PPN while, 
conversely,
the PPN describes effects outside of the 
pure-gravity sector of the SME.
The relationship between the PPN and other sectors of 
the SME is not known at present.

Many of the dominant effects in the pure-gravity sector of
the SME can be studied via an effective post-newtonian lagrangian for 
a system of point masses given by
\bea
L 
\hskip -4pt
&=&
\frac 12 \sum_{a} m_a \vec v^2_a 
+ \frac 12 \sum_{ab} \fr {G m_a m_b} {r_{ab}} (1+\frac 32 \sb^{00} 
+ \frac 12 \sb^{jk} \hat r^j_{ab} \hat r^k_{ab} )
\nonumber\\
&&
-\frac 12 \sum_{ab} \fr {G m_a m_b }{r_{ab}}( 3 \sb^{0j} v^j_a 
+ \sb^{0j} \hat r^j_{ab} v^k_a \hat r^k_{ab})
+ \ldots ,
\label{lpp}
\eea
where $v_a$ is the velocity of mass $m_a$ 
and $r_{ab}$ is the relative euclidean distance
between two masses.
In \rf{lpp}, 
the nine coefficients for Lorentz violation $\sb^\mn$ 
are projected into their space and time 
components $\sb^{00}=\sb^{jj}$, $\sb^{jk}$, 
and $\sb^{0j}$.
For other applications, 
such as the time-delay effect, 
it is necessary to directly use the post-newtonian metric 
which was obtained in \cite[Bailey \& Kosteleck\'y (2006)]{bk06}.

\section{Matter-gravity tests}

The chief effects from the coefficients $\ab_\mu$ 
can be described as an effective modification to the 
gravitational force between two test bodies.
Supposing there is a source body $S$, such as the Earth, 
and a test body $T$ near the surface, 
an addition to the usual vertical force arises
that takes the form
\beq
\widetilde F_z = 
- 2 g (\al \oaut_t + \al \oaus_t \mt/\ms),
\label{Fz}
\eeq
in a laboratory reference frame.
In this expression $\mt$ and $\ms$ 
are the masses of the test body and source, 
respectively, 
and $g$ is the local gravitational acceleration.
The force modification in \rf{Fz} depends on the time components 
of the $\ab_\mu$ coefficients for the test body and source, 
denoted $\oaut_t$ and $\oaus_t$.
For ordinary matter, 
these coefficients are ultimately related to 
the $12$ $\ab_\mu$ coefficients for the electron, neutron, 
and proton (denoted $\ab^e_\mu$, $\ab^n_\mu$, and $\ab^p_\mu$).

Two types of effects arise from the signal in \rf{Fz}.
When the coefficients ${\overline a}_\mu$ 
take on a flavor dependence,
violations of the Weak Equivalence Principle (WEP) occur, 
as two test bodies of different compositions
experience different accelerations.
The standard reference frame for reporting measurements
of coefficients for Lorentz violation is the 
Sun-centered celestial equatorial reference frame
or SCF for short (\cite[Kosteleck\'y \& Mewes 2002]{km02}).
When relating the laboratory frame coefficients 
in \rf{Fz} to the SCF, 
sidereal day and yearly time dependence 
from the Earth's rotation and revolution is introduced.
Therefore, 
an additional effect occurs where the signal for Lorentz violation 
depends on the time of day and season.
These effects have direct experimental consequences
for both Earth-bound and space-based tests of WEP
(\cite[Nobili et al.\ 2003]{gg03}), 
as well as ordinary single-flavor gravimeter-type tests.
Estimated sensitivities for specific tests and a single constraint
on one of the $12$ coefficients $\ab_\mu$ for ordinary matter,
implied by existing analysis (\cite[Schlamminger et al.\ 2008]{tp08}), 
are described in more detail in 
\cite[Kosteleck\'y \& Tasson (2009)]{kt09}. 

\section{Pure-gravity sector tests}

Within the pure-gravity sector of the SME, 
the primary effects on orbital dynamics, 
due to the nine coefficients $\sb^\mn$, 
can be obtained from the point-mass lagrangian in equation \rf{lpp}.
From the post-newtonian metric, 
modifications to the classic solar-system
tests of GR can be determined, 
such as the gyroscope experiment and 
the time-delay effect.

One particularly sensitive test of orbital 
dynamics in the solar system is lunar laser ranging.
Highly-sensitive laser pulse timing is achievable
using the reflectors on the lunar surface.
The dominant Lorentz-violating corrections 
to the Earth-Moon coordinate acceleration can be calculated 
from equation \rf{lpp}.
In the ideal case, 
a full analysis would incorporate
the effects from the pure-gravity sector of the minimal SME,
as well as the standard dynamics of the Earth-Moon system,
into the orbital determination program.
However, 
it can be shown that the dominant effects 
can be described as oscillations in the Earth-Moon distance.
The frequencies of these oscillations involve the mean orbital 
and anomalistic frequencies of the lunar orbit, 
and the mean orbital frequency of the Earth-Moon system
(see Table 2 in \cite[Bailey \& Kosteleck\'y 2006]{bk06}).
Using past data spanning over three decades, 
\cite[Battat et al.\ (2007)]{b07} placed 
constraints on $6$ combinations of the $\sb^\mn$ coefficients
at levels of $10^{-7}$ to $10^{-10}$.
Substantial improvement of these sensitivities
may be achievable in the future with APOLLO 
(\cite[Murphy et al.\ 2008]{m08}).
Also of potential interest are Earth-satellite tests, 
since satellites of differing orientation can
pick up sensitivities to distinct coefficients.

For Earth-laboratory experiments, 
a modified local gravitational acceleration arises
due to the $\sb^\mn$ coefficients, 
similar to that which occurred in \rf{Fz}.
Again, 
due to the Earth's rotation and revolution relative 
to the SCF, 
this acceleration acquires
a time variation that can be searched
for in appropriate experiments, 
such as gravimeter tests.
Such an experiment was performed by 
\cite[M\"uller et al.\ (2008)]{hm08}
and, 
more recently, 
these results were combined with lunar laser ranging
analysis to yield new constraints
on $8$ out of the $9$ independent $\sb^\mn$
coefficients (\cite[Chung et al.\ 2009]{c09}).

The classic time delay effect
of GR becomes modified in 
the presence of Lorentz violation.
The correction to the light travel time for a photon 
passing near a mass $M$ can be obtained
by studying light propagation with 
the post-newtonian metric of the pure-gravity sector 
of the minimal SME. 
If the signal is transmitted from 
an observer at position $r_e$, 
reflected from a planet or spacecraft at position $r_p$, 
the time delay of light can be written 
in a special coordinate system as
\beq
\De T_g \approx 4 GM \left[ (1+\sb^{TT}) 
\ln \left( \fr {r_e + r_p +R}{r_e +r_p -R} \right)
+\sb^{JK} \hat b^J \hat b^K \right].
\label{td}
\eeq
In this expression, 
$R$ is the distance between the observer and 
the planet or spacecraft and $\hat b$ is 
the impact parameter unit vector.
The coefficients for Lorentz violation 
are expressed in the SCF,
as denoted by capital letters.
This result holds for near-conjunction times when the 
photon passes near the mass $M$.
As discussed in \cite[Bailey (2009)]{b09}, 
time-delay tests can be useful to constrain
the isotropic $\sb^{TT}$ coefficient, 
among others.
It would be of definite interest to perform data analysis
searching for evidence of nonzero $\sb^\mn$ coefficients 
in sensitive time-delay experiments such as 
Cassini (\cite[Bertotti et al.\ 2003]{cass03}), 
BepiColombo (\cite[Iess \& Asmar 2007]{ia07}),
and SAGAS (\cite[Wolf et al.\ 2009]{sagas}). 

In addition to effects discussed above, 
many standard results of GR receive
modifications in the presence of Lorentz violation.
This includes classic tests
such as the perihelion shifts of the inner planets, 
the gravitational redshift,
and the classic gyroscope experiment.
In addition, 
tests beyond the solar system, 
such as with binary pulsars (\cite[Kramer 2009]{kramer09} 
and \cite[Stairs 2009]{stairs09}), 
can also probe SME coefficients.
In the case of binary pulsars, 
the dominant effects arise from changes 
in the orbital elements of the pulsar-companion 
oscillating ellipse. 
For more details the reader is referred to 
\cite[Bailey (2009)]{b09} and 
\cite[Bailey \& Kosteleck\'y (2006)]{bk06}.

\section{Acknowledgments}

Q.G.\ Bailey wishes to thank the International Astronomical Union
for the travel grant that was provided to attend the symposium.

\end{document}